\documentclass[twocolumn,english,aps,prb]{revtex4}
\usepackage[T1]{fontenc}
\usepackage[latin9]{inputenc}
\usepackage{color}
\usepackage{amsmath}
\usepackage{graphicx}
\usepackage{amssymb}
\usepackage{esint}
\usepackage{wasysym}

\makeatletter
\@ifundefined{textcolor}{}
{%
 \definecolor{BLACK}{gray}{0}
 \definecolor{WHITE}{gray}{1}
 \definecolor{RED}{rgb}{1,0,0}
 \definecolor{GREEN}{rgb}{0,1,0}
 \definecolor{BLUE}{rgb}{0,0,1}
 \definecolor{CYAN}{cmyk}{1,0,0,0}
 \definecolor{MAGENTA}{cmyk}{0,1,0,0}
 \definecolor{YELLOW}{cmyk}{0,0,1,0}
 }

\@ifundefined{definecolor}
 {\usepackage{color}}{}
\@ifundefined{definecolor}{\@ifundefined{definecolor}
 {\usepackage{color}}{}
}{}
\usepackage{epsf}

\newcommand{\nio}{Na$_2$IrO$_3$}
\newcommand{\lio}{Li$_2$IrO$_3$}
\newcommand{\aio}{A$_2$IrO$_3$}

\newcommand{\xis}{\xi_{\rm S}}

\newcommand{\xiSG}{\xi_{\rm SG}}
\newcommand{\xp}{x_p}

\newcommand{\TN}{T_{\rm N}}
\newcommand{\Tg}{T_{\rm g}}
\newcommand{\Tu}{T_{\rm u}}
\newcommand{\Tl}{T_{\rm l}}

\newcommand{\Jp}{J_{\perp}}
\newcommand{\jt}{$J_1$-$J_2$-$J_3$ }
\newcommand{\hk}{Heisenberg-Kitaev }
\newcommand{\Nr}{N_{\rm rl}}

\makeatother

\usepackage{babel}

\begin{document}

\title{
Magnetism in spin models for depleted honeycomb-lattice iridates:\\Spin-glass order towards percolation
}

\author{Eric C. Andrade}
\affiliation{Institut f\"ur Theoretische Physik, Technische Universit\"at Dresden,
01062 Dresden, Germany}
\affiliation{Instituto de F\'isica Te\'orica, Universidade Estadual Paulista, Rua
Dr. Bento Teobaldo Ferraz, 271 - Bl. II, 01140-070, S\~ao Paulo, SP, Brazil}
\author{Matthias Vojta}
\affiliation{Institut f\"ur Theoretische Physik, Technische Universit\"at Dresden,
01062 Dresden, Germany}


\begin{abstract}
Iridates are characterized by a fascinating interplay of spin-orbit and electron-electron
interactions. The honeycomb-lattice materials A$_2$IrO$_3$ (A=Na,Li) have been proposed to
realize pseudospin-1/2 Mott insulating states with strongly anisotropic exchange
interactions, described by the Heisenberg-Kitaev model, but other scenarios involving
longer-range exchange interactions or more delocalized electrons
have been put forward as well.
Here we study the influence of non-magnetic doping, i.e., depleted moments, on the magnetic
properties of experimentally relevant variants of the Heisenberg-Kitaev and Heisenberg
$J_1$-$J_2$-$J_3$ models. We generically find that the zigzag order of the clean system
is replaced,  upon doping, by a spin-glass state with short-ranged zigzag correlations. We
determine the spin-glass temperature as function of the doping level and show that
this quantity allows to assess the importance of longer-range exchange interactions
when the doping is driven across the site percolation threshold of the honeycomb lattice.
\end{abstract}

\date{\today}


\maketitle


\section{Introduction}

The interplay of strong spin-orbit coupling and electronic correlations is at the heart
of many recent developments in condensed-matter physics, involving, e.g., correlated
topological insulators, fractional Chern insulators, and spin-orbit Mott insulators
\cite{pesin10,hohenadler13,bergholtz13,kim08,kim09}. On the materials side, oxides with
partially filled 5d shells, such as iridates and osmates, are considered promising
candidates in order to realize the theoretically proposed phenomena.

In this context, the insulating iridates \aio\ (A=Na,Li) have attracted enormous
attention over the past few years \cite{Sin10,Liu11,Choi12,Sin12,Ye12,Com12}. In these
materials, the Ir$^{4+}$ ions are arranged in a layered honeycomb-lattice structure. Due
to the combined effect of strong spin-orbit coupling and Coulomb interactions, the Ir
5d$^{5}$ states, with one hole in the t$_{2g}$ manifold, have been proposed to realize
$J_{{\rm eff}}=1/2$ spin-orbit Mott insulators \cite{Shi09,Jac09}, similar to other
layered iridates \cite{kim08,kim09}. Furthermore, Ref.~\onlinecite{Cha10} suggested that
the magnetism of the $J_{{\rm eff}}=1/2$ moments is dominated by strongly
spin-anisotropic compass interactions, which by itself lead to the spin-liquid model on
the honeycomb lattice proposed by Kitaev \cite{Kit06}. Supplemented by an additional
spin-isotropic Heisenberg interactions, the resulting \hk (HK) model has been shown to host
both spin-liquid and conventionally ordered phases
\cite{Cha10,Jia11,Reu11,Bha12,Cha13,perkins12,perkins13}.

Experimentally, both {\nio} and {\lio} have been found to undergo
a magnetic ordering transition at $\TN\simeq15$\,K \cite{Sin10,Liu11,Sin12}.
In {\nio} the low-temperature spin configuration has been identified
as collinear ``zigzag'' order \cite{Choi12,Ye12}, with ferromagnetic
zigzag chains arranged antiferromagnetically in the honeycomb plane.
This state is indeed a ground state of the HK model, where it results
from a competition of antiferromagnetic Kitaev and ferromagnetic Heisenberg
interactions \cite{Cha13}. Alternatively, Heisenberg and HK models
with longer-range interactions have been considered: specifically,
a Heisenberg \jt model with sizeable second and third-neighbor coupling
has been found to describe the available data as well \cite{Kimchi11,Choi12}.
Finally, a more itinerant scenario in terms of molecular orbitals
has also been proposed \cite{Maz12}, although a detailed description
of the magnetic properties in this model is lacking to date.

\begin{figure}[t]
\includegraphics[width=0.48\textwidth]{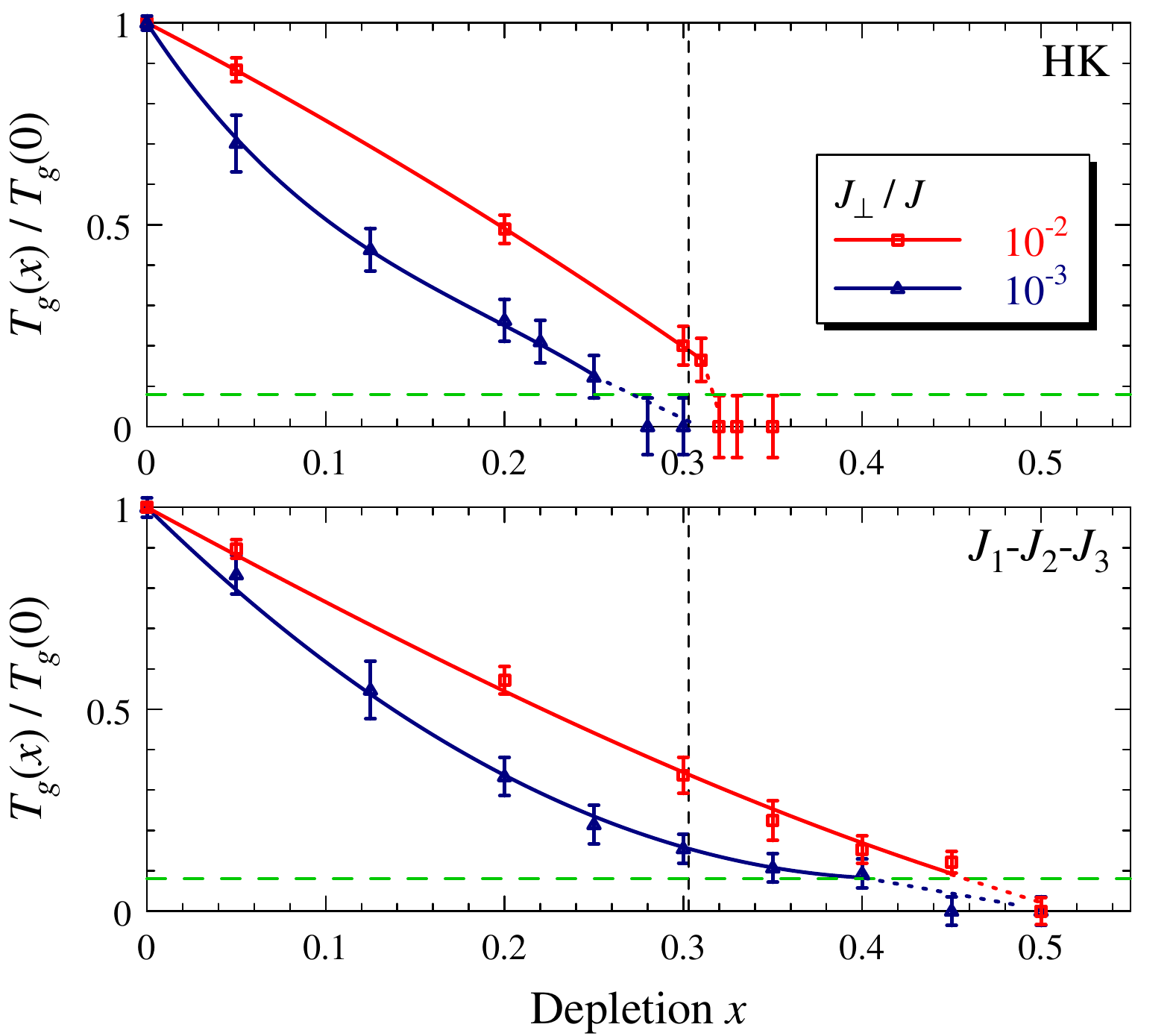}
\caption{
\label{fig:tg}
Ordering/freezing temperature extracted from our MC simulations, shown as $\Tg(x)/\Tg(x=0)$ as function of doping level $x$, for the HK (top) and \jt (bottom) models for two different values of the interlayer coupling $\Jp/J$. The vertical dashed line locates the 2d
percolation threshold $\xp$; the horizontal dashed line marks temperatures below which we are
unable to reach equilibrium in our MC simulations.
Solid lines are polynomial fits through the data; dotted lines are extrapolations.
}
\end{figure}

In this paper, we propose magnetic depletion, i.e., the random substitution of magnetic
Ir$^{4+}$ by non-magnetic ions, as a powerful tool to study the magnetism of {\aio} and
to discriminate between the various proposed scenarios for magnetism. A key insight is
that, within local-moment models, depletion will inevitably turn the zigzag ordered state
into a spin (or spin-orbit) glass: Both the HK and \jt models are frustrated, and the
combination of disorder and frustration generically causes spin-glass behavior
\cite{villain79}. We calculate the freezing temperature, $\Tg(x)$, as function of doping
level $x$ and show that its behavior across the site-percolation threshold, $\xp=30.3\%$,
strongly differs between the HK and \jt models, Fig.~\ref{fig:tg}.

\section{Models}
%
We focus on two models which have been considered to describe the zigzag ordered state of
\nio. The first is the HK model,
\begin{equation}
\mathcal{H}= J \sum_{\left\langle ij\right\rangle} \vec{S}_{i}\cdot\vec{S}_{j} + 2K \sum_{\left\langle ij\right\rangle
_{\gamma}}S_{i}^{\gamma}S_{j}^{\gamma},
\label{hk}
\end{equation}
the second the \jt model
\begin{equation}
\mathcal{H}=J_{1}\sum_{\left\langle ij\right\rangle }\vec{S}_{i}\cdot\vec{S}_{j}+J_{2}\sum_{\left\langle \left\langle ik\right\rangle \right\rangle }\vec{S}_{i}\cdot\vec{S}_{k}+J_{3}\sum_{\left\langle \left\langle \left\langle il\right\rangle \right\rangle \right\rangle }\vec{S}_{i}\cdot\vec{S}_{l}.
\label{jt}
\end{equation}
Here, the sums run over pairs of nearest, second, and third neighbor sites, respectively,
while $\gamma=x$, $y$, $z$ in Eq.~\eqref{hk} labels the three different links for each
spin in a honeycomb lattice.
The parameter regimes of interest are defined through the presence of zigzag magnetic
order as realized in {\nio} \cite{Choi12,Ye12}.
The HK model's couplings may be parametrized as $J=A\cos\phi$ and
$K=A\sin\phi$, where $A$ is an overall energy scale. Its full
phase diagram was first mapped out in Ref.~\onlinecite{Cha13}, with the zigzag phase
occurring for $0.51\pi<\phi<0.90\pi$;
in the following we choose $\phi=0.62\pi$.
For the \jt model, sizeable $J_{2}$ and $J_{3}$ are required in order to have a
zigzag magnetic ground state \cite{rastelli79,lhui01,li12}. Following
Ref.~\onlinecite{Choi12}, we choose $J_{2}=0.8J_{1}$ and $J_{3}=0.9J_{1}$.

As will become clear below, the magnetic properties of the depleted
HK and \jt models depend sensitively on the presence of a magnetic
coupling between the layers. For \aio, no quantitative information
on such coupling is available at present; it is often assumed to be
small due to the A-B-type stacking of the honeycomb layers. Here we
will account for the 3d character by considering a layered model with
A-A stacking and a small vertical (unfrustrated) Heisenberg coupling $\Jp$;
in application to \aio\ this is to be understood as an effective
coupling between second-neighbor layers.


\section{Monte-Carlo simulations}
%
We study the models \eqref{hk} and \eqref{jt} using classical Monte Carlo (MC)
simulations for unit-length spins on lattices of size $L\times L\times L_{z}$, typically
with $L_{z}=L/2$ and periodic boundary conditions. The honeycomb layers are spanned by
the primitive lattice vectors $\vec{a}_{1\left(2\right)}=\left(3/2,\pm\sqrt{3}/2\right)$,
with each unit cell containing two sites. Depletion is simulated by randomly removing a
fraction $x$ of spins, with $x$ varying between 5\% and 50\%, with the total number of
spins $N_{s}=\left(1-x\right)\times2L^{2}L_{z}$.
We perform equilibrium MC simulations using single-site updates with a combination of the
heat-bath and microcanonical (or over-relaxation) methods, with typically
$10^{6}$ MC steps per spin, and combine this with the parallel-tempering algorithm \cite{mcdetails,av12a}. Disorder averages are taken over $\Nr$
samples, with $\Nr$ ranging from $1000$ for $L=8$ to $\Nr=50$ for $L=20$.
Below we quote energies in units of $J\equiv J_{1}$, the nearest-neighbor Heisenberg exchange.

We extract the ordering (or freezing) temperature $\Tg$ from the crossing points of
$\xi(T)/L$ for different $L$, according to the scaling law $\xi/L=f(L^{1/\nu}(T-\Tg))$,
where $\xi$ is a correlation length, $f(x)$ a scaling function, and $\nu$ the correlation
length exponent. This procedure is especially suitable to detect spin-glass freezing, as
shown in previous studies of the 3d Edwards-Anderson model
\cite{lee_young1,campos06,viet09}. The main source of numerical error in $\Tg$ is from
the $L\to\infty$ extrapolation of the crossing point location required for small $L$.

The magnetic correlation length $\xis$ is calculated from a fit of the static magnetic
structure factor, $S(\vec{q})$, close to the ordering wavevector $\vec{Q}$ (the three
independent $\vec{Q}$ vectors corresponding to the zigzag order are
$(\vec{b}_{1}+\vec{b}_{2})/2$, $\vec{b}_{1}/2$, and $\vec{b}_{2}/2$, where
$\vec{b}_{1(2)}=2\pi(1/3,\pm1/\sqrt{3})$ are the
reciprocal lattice vectors).
Analogously, the spin-glass correlation length $\xiSG$ is obtained
from the spin-glass susceptibility
$\chi_{SG}(\vec{q})=N_{s}\sum_{\alpha,\beta}\big[\left\langle
\left|q^{\alpha,\beta}\left(\vec{q}\right)\right|^{2}\right\rangle \big]_{av}$, where
$q^{\alpha,\beta}\left(\vec{q}\right)=N_{s}^{-1}\sum_{i}S_{i}^{\alpha\left(1\right)}S_{i}^{\beta\left(2\right)}\mbox{exp}\left(i\vec{q}\cdot\vec{r}_{i}\right)$
is the spin-glass order parameter. Here $\alpha$ and $\beta$ are spin components,
$^{(1,2)}$ denote identical copies of the system (``replicas'') containing
the disorder configuration, $\langle\cdots\rangle $ denotes MC average, and
$[\cdots]_{av}$ average over disorder.


\section{Clean HK model}
%
The 2d disorder-free HK model has been
studied by various numerical methods \cite{Cha10,Cha13,Reu11,perkins12,perkins13}.
A comparison of phase diagrams shows that the classical-spin HK
model reproduces \cite{perkins13} all phases of the spin-1/2 model
except for the quantum spin liquid \cite{Cha13}, with $T=0$ phase boundary locations
in reasonable agreement between quantum and classical models. The
results in Refs.~\onlinecite{perkins12,perkins13} also indicate
two thermal transitions upon cooling
to any of the ordered low-$T$ phases.
The system enters a critical phase at
$\Tu$, with power-law spin correlations, and a state with
true long-range order is reached only below $\Tl < \Tu$. This behavior
parallels that of a 2d six-state clock model \cite{clock}, as suggested by the
sixfold degeneracy of the ordered states in the HK model.

\begin{figure}[t]
\includegraphics[width=0.49\textwidth]{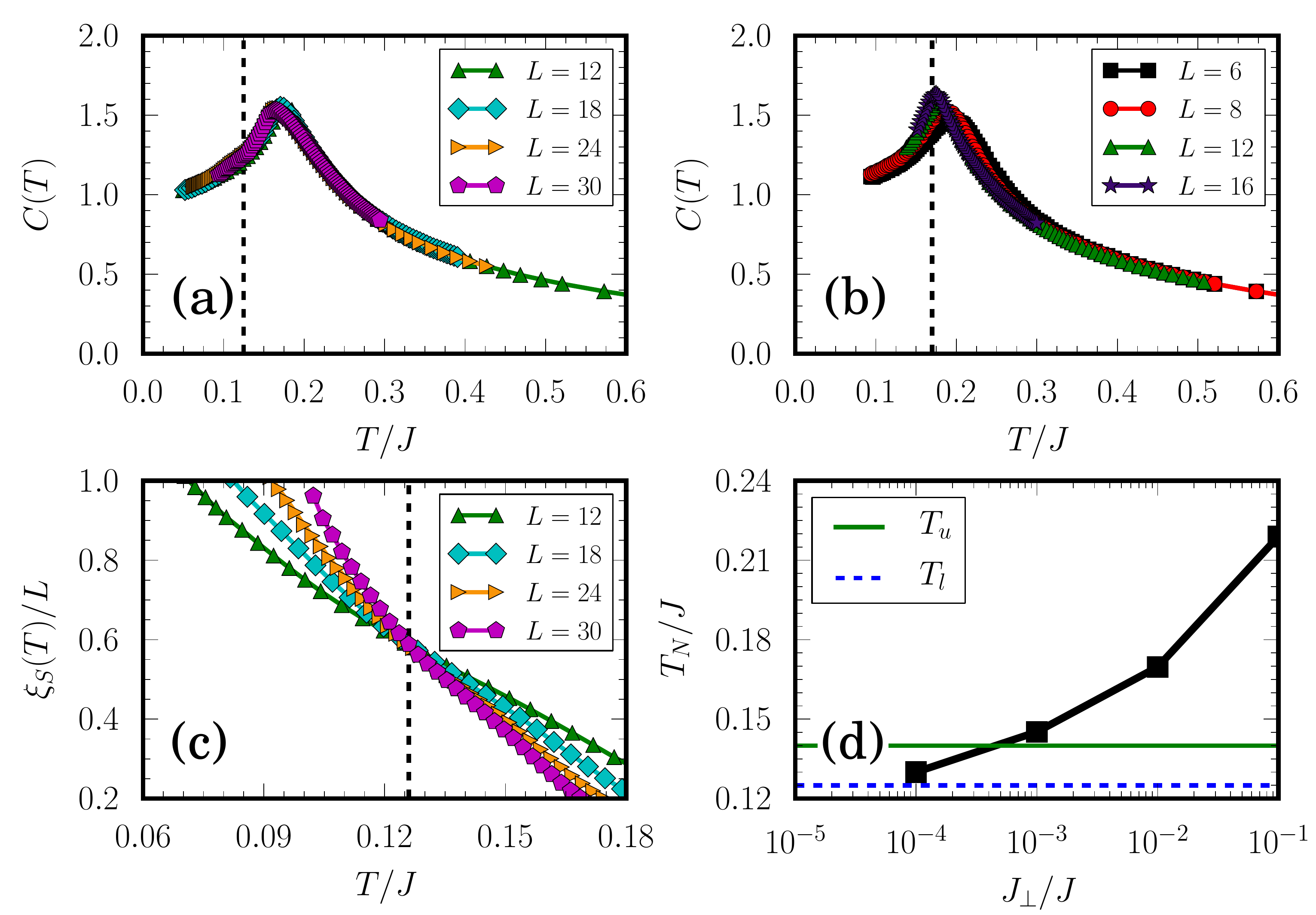}
\caption{
\label{fig:clean}
MC results for the ordering into the zigzag state
of the clean HK model for several system sizes $L$.
(a,c): 2d. (b,d): 3d.
(a,b): Specific heat as a function of the temperature $T$, in (b) with $\Jp/J=10^{-2}$.
(c): $\xis/L$ as a function of $T$. The vertical dashed line indicates $\Tl$.
(d):  $\TN(\Jp/J)$, also showing $\Tu$ and $\Tl$ of the 2d HK model;
the value of $\Tu$ was extracted from Ref.~\onlinecite{perkins13}.
}
\end{figure}

For selected values of $\phi$, we have verified that our MC simulations,
applied to the 2d HK model ($\Jp=0$), reproduce the results of
Ref.~\onlinecite{perkins13}. In particular, the specific heat,
Fig.~\ref{fig:clean}(a), shows a broad peak far above both $\Tu$
and $\Tl$ while the singularity at the transitions is weak
-- this reflects the presence of strong fluctuations
in the 2d system. Nevertheless, there is a well-defined crossing point
in $\xis/L$ at $\Tl$ where long-range order sets in, Fig.~\ref{fig:clean}(c).

We have then switched on the inter-layer coupling $\Jp$ and monitored
the evolution of the transition temperature, Fig.~\ref{fig:clean}(d).
As expected on general grounds, the critical intermediate phase of
the 2d system disappears for finite $\Jp$, such that there is only
a single thermal phase transition at $\TN$, which now displays a
pronounced specific-heat singularity, Fig.~\ref{fig:clean}(b). For
$\Jp/J\gtrsim10^{-3}$, $\TN$ is larger than both $\Tl$ and $\Tu$
of the 2d system, and our data is compatible with $\TN\to\Tl$ as $\Jp\to0$,
although finite-size effects hamper an accurate determination of $\TN$
for $\Jp/J<10^{-4}$, Fig.~\ref{fig:clean}(d).

\textit{Clean \jt model.}
We have also performed corresponding simulations
for the \jt model. Here, $\TN\to0$ for $\Jp\to0$ due to the assumed
continuous spin symmetry.
For $\Jp/J=10^{-2}\left[10^{-3}\right]$ we have $\TN/J=0.446\left(3\right)\left[0.42\left(1\right)\right]$.

\section{Magnetic depletion}
%
We now describe our central results,
obtained for the depleted HK and \jt models, with a concentration
$x$ of randomly placed vacancies. Since both models are frustrated,
the introduction of vacancies generate local non-collinearities in
the spin order \cite{henley89}, which ultimately leads to spin-glass
behavior \cite{villain79}.
\begin{figure}[t]
\includegraphics[width=0.48\textwidth]{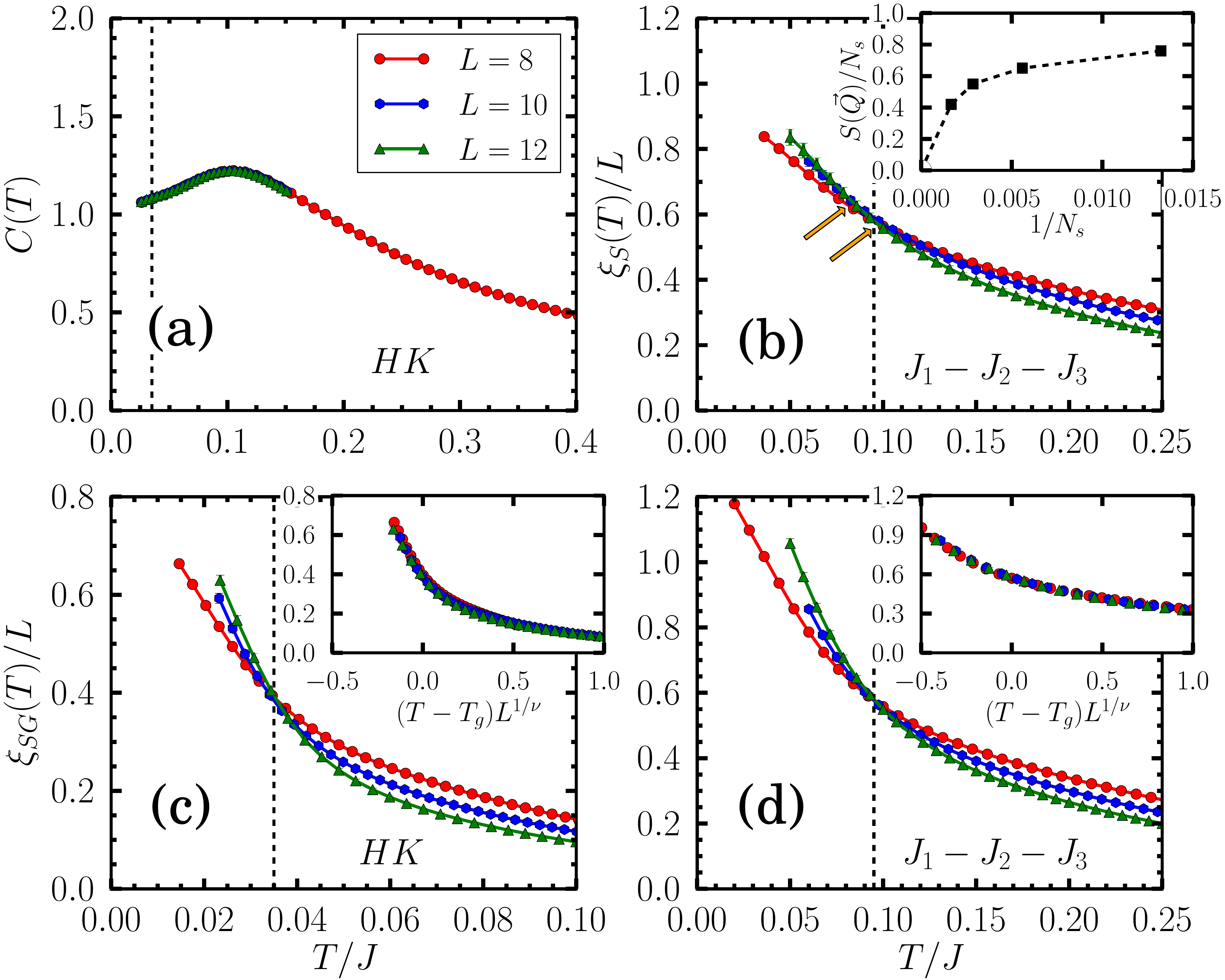}
\caption{
\label{fig:dirty}
MC results for the ordering into the spin-glass state for $\Jp/J=10^{-2}$.
(a,c): HK model at $x=30\%$. (b,d): \jt model at $x=35\%$.
(a) Specific heat as a function of the temperature $T$.
(b) $\xis/L$ as a function of $T$. The arrows indicate the crossing points of the curves
for different pairs of $L$, displaying a clear downward trend with increasing $L$. Inset:
Bragg intensity $S(\vec{Q}/N_{s})$ extrapolated to $T\to 0$ as a function of $1/N_{s}$.
(c,d) $\xiSG/L$ as a function of $T$. Inset: Scaling plot with $\xiSG/L$ as a function of
$(T-\Tg)L^{1/\nu}$.
The vertical dashed lines indicate the glass temperature $\Tg$, as determined
in (c) and (d).
}
\end{figure}

We have first studied the 2d case ($\Jp=0$) and found -- in both models and for any $x\geq5\%$
-- indications of neither conventional nor spin-glass order at finite temperature. This
is expected: conventional order is suppressed, due to the combination of disorder and
frustration, in favor of spin-glass magnetism. However, the glass temperature is strictly
zero in two dimensions \cite{Bray84,fischer} even in the case of Ising symmetry.

For finite interlayer coupling the situation changes,
with sample results shown in Fig.~\ref{fig:dirty}. While conventional
long-range order is absent for any $x\geq 5\%$, spin-glass order emerges
instead at low $T$.
The latter is signified by a well-defined common crossing point in $\xiSG/L$ and a
corresponding scaling, Fig.~\ref{fig:dirty}(c)-(d) \cite{nu}.
In contrast, existing crossing points of $\xis/L$ display a systematic downward shift
with increasing $L$, indicative of short-range zigzag spin correlations,
Fig.~\ref{fig:dirty}(b). We note that we do not reach the limit $L\gg\xis(T=0)$ where crossing
points would be absent entirely.

Short-range magnetic order also manifests itself in the specific heat,
Fig.~\ref{fig:dirty}(a). The peak in $C(T)$ is broad and occurs at a temperature
considerably larger than the freezing temperature (here $T_{\rm peak}\approx2\Tg$),
indicating that this short-range order builds up at temperature considerably higher than
$\Tg$. We stress that this behavior is a hallmark of glassy systems \cite{av12a,fischer},
and it is, in principle, disconnected from the non-trivial behavior of the 2d
disorder-free HK model \cite{perkins12}, Fig.~\ref{fig:clean}(a).

To account for the possibility of different (non-zigzag) dilution-induced magnetic ground
state, we monitored $S(\vec{q})$ in the reciprocal space, but (within our resolution) we
detected peaks only at the $\vec{Q}$ vectors corresponding to the zigzag order. However,
these peaks grow slower than the system size, Fig.~\ref{fig:dirty}(b), again indicating
static short-range order with a vanishing magnetic order parameter.

%

\begin{figure}[t]
\includegraphics[width=0.49\textwidth]{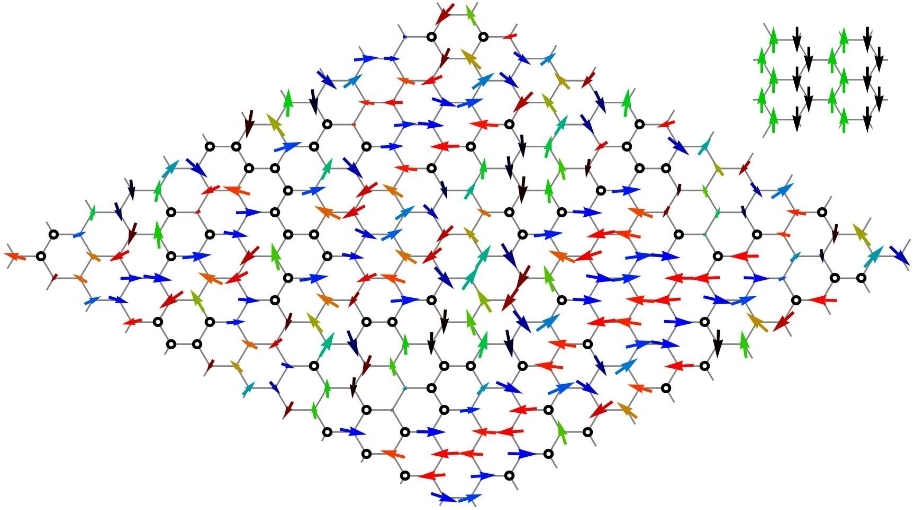}
\caption{\label{fig:spatial}
Sample ground-state spin configuration of the classical HK model at $x=20\%$ depletion,
with one $L=12$ layer shown. The arrows denote the $x$ and $z$ components of the
$\vec{S}_{i}$; the circles indicate the vacancy positions. The arrow lengths indicate the
weight of the projection onto the $x-z$ plane and the colors the in-plane orientation.
Short-range zigzag order with glassy domain formation is visible.
Inset: Ideal zigzag order with the spins aligned along $S_{i}^{z}$.
}
\end{figure}

\section{Small doping and glassiness}
%
In both the \hk and \jt models, we find a single vacancy to produce anticollinear states \cite{henley89}. Multiple vacancies have a somewhat different effect in the two models: In the \hk case, the vacancies locally select specific stripe orientations due to spin-orbit coupling \cite{trousselet11}, causing domains with different stripe orientation to coexist. In the \jt model, instead, the effect of long-range distortions of the spin pattern is more prominent, due to the presence of gapless bulk modes.
Remarkably, for the vacancy concentrations of interest, $x\ge 20\%$, the spin configurations we observe in both models are virtually indistinguishable and are characterized by a short-range domain structure, Fig. \ref{fig:spatial}.

Based on our MC data, we are unable to decide whether long-range order is destroyed in favor of spin-glass order at infinitesimal $x$ or at a finite critical $x_c$ (with $x_c<5\%$). We leave a more detailed characterization of the small-doping behavior for future work.


\section{Ordering temperature and percolation}
%
An easily accessible
quantity is the ordering (or freezing) temperature $\Tg$ as function
of $x$. While one generally expects that $\Tg$ decreases with increasing
$x$, the behavior at large $x$ contains information on the nature
of the magnetic couplings: For a layered local-moment system with
nearest-neighbor couplings, $\Tg$ will diminish near the threshold
$\xp$ for 2d site percolation, because for $x>\xp$ the layers fragment
into disconnected spin clusters, and for Heisenberg symmetry $\Tg(x)/\Tg(x=0)$
will vanish as $x\to\xp$ in the limit of small interlayer coupling.
In contrast, in systems with longer-range magnetic couplings, $\Tg$
will stay finite across $\xp$ \cite{percfoot}.
The parent compounds of cuprate superconductors
beautifully exemplify this physics: the Neel temperature in Zn-doped
La$_{2}$CuO$_{4}$ vanishes essentially at the square-lattice percolation
threshold of $\xp=40.5\%$ \cite{greven02} -- this proves that the
cuprate magnetism is dominated by nearest-neighbor coupling.

Our results for $\Tg(x)$ are shown in Fig.~\ref{fig:tg}. As expected from the above
discussion, $\Tg$ of the nearest-neighbor HK model rapidly drops towards the
honeycomb-lattice $\xp=30.3\%$ and becomes smaller than our lowest simulation temperature
($J/80$) for $x\gtrsim32\%$ for $J_\perp/J=10^{-2}$.
(Note that, due to the finite $\Jp$, $\Tg$ is expected to be non-vanishing up to the 3d
percolation threshold, however, it is undetectably small for $x\gtrsim32\%$.)
For smaller interlayer coupling this apparent vanishing of $\Tg(x)$ appears at even smaller $x$.

In contrast, $\Tg$ of the \jt model continues its approximately
linear variation with $x$ across $\xp$ and extrapolates to our lowest
simulation temperature at a much larger doping level of $x\approx50\%$ \cite{triang}.
For both models, $\Tg(x)/\Tg(x\!=\!0)$ diminishes with decreasing $J_\perp/J$, and small $J_\perp/J$ induce a curvature in $\Tg(x)$ which is particularly pronounced at small $x$.


\section{Summary}
%
We have studied the magnetism of local-moment models for {\aio} under magnetic depletion.
A spin-orbit glass, with zigzag short-range order, emerges generically from the
combination of strong spin-orbit coupling, frustration, and disorder. We have determined
the glass (or freezing) temperature $\Tg$ as function of the doping level $x$, which at
large doping differs qualitatively between the HK and \jt models, Fig.~\ref{fig:tg}(a).

We thus propose to employ magnetic depletion, using dopants with magnetically
inert $d$ shells, as a tool to assess the importance of
longer-range magnetic couplings in the {\aio} compounds: If the experimental $\Tg$ were
found to vanish near $\xp$ this would strongly hint \cite{itfoot} at short-range HK
physics being realized in {\aio}, as originally proposed in
Refs.~\onlinecite{Cha10,Cha13}. Conversely, the absence of such vanishing would imply significant longer-range interactions.

\textit{Note added.}
Very recent experiments \cite{geg14}, using non-magnetic Ti dopants substituting for Ir, show significant differences between depleted {\nio} and {\lio} across the percolation threshold.

\acknowledgments

We thank J. van den Brink, P. Gegenwart, P. Horsch, G. Khaliullin, and A. Rosch
for discussions. The computations were partially performed 
at the Center for Information Services and High Performance Computing
(ZIH) at TU Dresden.
This research was supported by the DFG through FOR 960 and GRK 1621
as well as by the Helmholtz association through VI-521. E.C.A. was also
partially supported by FAPESP.



\begin{thebibliography}{99}
\bibitem{nu}
For both models, we estimate the same correlation length exponent $\nu\approx1.05$. Due to our modest system sizes, we do not pursue a complete characterization of the critical exponents.
\bibitem{percfoot}
A lattice with further-neighbors links has a larger site percolation threshold than the one with nearest-neighbor links only.
\bibitem{itfoot}
In more itinerant scenarios of {\aio} magnetism, such as in Ref.~\onlinecite{Maz12}, site
percolation is not expected to drastically influence $\Tg(x)$.
\end{thebibliography}
\end{document}